# The Alex Catalogue, A Collection of Digital Texts with Automatic Methods for Acquisition and Cataloging, User-Defined Typography, Cross-searching of Indexed Content, and a Sense of Community

*Eric Lease Morgan*

Digital Library Initiatives
Box 7111, NCSU Libraries, Raleigh, NC 27695-7111, USA
Tel: (919) 515-4221
E-mail: eric_morgan@ncsu.edu

**ABSTRACT**
This paper describes the Alex Catalogue of Electronic Texts, the only Internet-accessible collection of digital documents allowing the user to 1) dynamically create customized, typographically readable documents on demand, 2) search the content of one or more documents from the collection simultaneously, 3) create sets of documents from the collection for review and annotation, and 4) publish these sets of annotated documents in turn fostering a sense of community around the Catalogue. [1] More than a just a collection of links that will break over time, Alex is an archive of electronic texts providing unprecedented access to its content and features allowing it to meet the needs of a wide variety of users and settings. Furthermore, the process of maintaining the Catalogue is streamlined with tools for automatic acquisition and cataloging making it possible to sustain the service with a minimum of personnel.

**HISTORY**
The original Alex Catalogue of Electronic Texts was announced as a gopher services by its creator, Hunter Monroe, in 1994. [2] At that time, the Catalogue was a set of gopher link files; no content was archived locally. The link files were the result of reports generated against a DBase II application. It was soon after the announcement was made when the author and Monroe became acquainted. It was not long after that when it was decided to mirror the Alex Catalogue on the Gopher at the NCSU Libraries within its Library Without Walls.

In 1994, the combined use of Mosaic and the hypertext transfer protocol (HTTP) were well along their way to becoming the preferred means for "surfing the Web". Monroe was encouraged by the author to generate HTML files from his DBase II application. Monroe obliged and Alex became available via HTTP. By creating a highly structured report this new version of the Catalogue was indexed and made searchable via a Wide Area Information Server (WAIS) and a WAIS/HTTP gateway (freewais-sf and SFgate, respectively). In an effort to begin fulfilling the ideas set forth in Monroe's original announcement, he was asked to generate one more report, a report that could be used to create MARC records. This report was put through a locally developed utility generating MARC records. Using scripts originally written by Tim Kambitsch, the MARC records were uploaded into one of the Internet's first Web-accessible OPAC databases. It was called Alcuin. [3]

For about two years Alex and its derivatives languished in neglect. Of the original 1,200 links many broke over time. Despite this fact, the original Alex implementations continue to be referenced and used. Considering Alex's popularity it was decided to move the site to a newer, more robust architecture in early 1998. This new architecture provides better readable, browsable, and searchable features. More importantly, this new architecture reduces the possibility of broken links by archiving the texts in the collection.

The balance of this paper describes the purpose and scope, features, technical infrastructure, discussion, and future directions of the newest implementation of Alex.

**PURPOSE AND SCOPE**
The purpose of the Alex Catalogue is two-fold. Its primary purpose is to assist the author in demonstrating a concept coined "arscience", a process of understanding using methods from both art and science. Arscience is an epistemological method employing a Hegalian dialectic juxtaposing approaches to understanding that include art and science, synthesis and analysis, as well as faith and experience. These epistemological methods can be compared and contrasted in passages of great literature. The literature in the Catalogue can be of assistance in demonstrating this concept.

The Catalogue's secondary purpose is to provide value-added access to some of the world's great literature. Great literature is broadly defined as literature withstanding the test of time, literature expressing time-tested, universal human values. The process of reading great literature exposes people to these values and provides opportunities for learning and the acquisition of knowledge. A purpose of the Alex Catalogue is to increase the number of these opportunities.

The Alex Catalogue of Electronic Texts is a collection of digital documents. The scope of the documents in the collection is limited to primary materials from American and English literature as well as Western philosophy. Listed in priority order, texts in the collection must have the following qualities:

1. Only texts in the public domain or freely distributed texts are collected.
2. Only texts that can be classified as American literature, English literature, or Western philosophy are included.
3. Only texts that are considered "great" literature are included. Great literature is defined as literature withstanding the test of time and found in authoritative reference works like the Oxford Companions or the Norton Anthologies. Items found in these sorts of reference works usually exemplify items articulating universal human values.
4. Only complete works are collected unless a particular work was never completed in the first place; partially digitized texts are not be included in the Catalogue.
5. Whenever possible, collections of short stories or poetry are be included as they were originally published. If the items from the originally published collections have been broken up into individual stories or poems, then those items will be included individually.
6. The texts in the collection must be written in or translated into English. Otherwise the author will not be able to evaluate the texts' quality nor will the indexing and content-searching work correctly.

Because of technical limitations and the potential long-term integrity of the Catalogue, texts in the collection, listed in order of preference, have the following formats:

1. plain text files are preferred over HTML files
2. HTML files are preferred over compressed files
3. compressed files are preferred over "word processor" files
4. word processed files are the least preferable file format
5. texts in unalterable file formats, such as Adobe's Portable Document Format (PDF), will not be included

In all cases, whole texts are preferred over texts that have been divided. If a particular text is deemed especially valuable and the text has been divided into parts, then efforts will be made to concatenate the individual parts and incorporate the result into the collection. The items in the collection are not necessarily intended to be read online.

**FEATURES**
The Alex Catalogue of Electronic Texts has a number of features used to satisfy the purposes outlined above.

**Browsability and Searchability**
Initial access to the documents of the Catalogue is provided through a browsable as well as a searchable interface. The browsable interface is accessible via authors, titles, and/or filenames. The author browsable list enumerates all the authors in the collection's author authority table. Each enumeration is marked up in such a way that selecting any one of the authors invokes a search against the Catalogue for that author and results are returned accordingly. The browsable title list works in a similar manner. Each letter in the alphabet is listed on an HTML page. By selecting an item from the list a title search is initiated and results are returned. The browsable filename list is admittedly the least useful of the Catalogue's access methods but since the physical items of the collection are stored to the computer's file system according to centuries, the browsable filename lists provide rudimentary access according to date.

The searchable interface's input options support free text and field queries against the metadata (catalog records) describing items in the collection. These queries can include Boolean operations, exact matches, phrase searches, as well as a stemming function. Searches are case-insensitive by default but can be selected otherwise. The search engine does not support nested queries but it does support a kind of saved sets operation as described in the "bookcases" section below.

The searchable interface's output options are limited to three options: 1) titles, 2) titles, authors, and links, and 3) full records. The titles option returns just the titles of located texts. These titles are hyperlinked to the original location where they were found on the Internet.

The titles, authors, and links option is the default and provides access to the original location of the document as well as the locally archived location. This option also provides links allowing the user to create a typographically readable document of the located texts (described below). The titles, authors, and links option also provides links allowing the user to search content of one or more of the located documents (described below).

The final output option, full records, is a super-set of option #2. Each full record includes all the information from option #2, the document's physical description as well as subject and genre assignments. Presently, no records in the Catalogue have subject or genre assignments, and the full record display states this fact.

**Typographically Readable Documents**

A unique feature of the Catalogue is the on-the-fly creation of PDF files, typographically readable documents on demand. This feature allows the user to create documents, intended for printing, that are more readable than plain text files with hard carriage returns. Using this option the user can specify things like fonts and font sizes for their output. Older people who may have visual impediments might choose to create their documents using a large san serif font like Helvetica 16 point. On the other hand, a younger person may want to have their documents produced using something more traditional like Times 12 point. The creation of these PDF documents allows the user to create simply formatted but very readable documents for printing. The documents in the collection are not necessarily intended to be read online.

**Content Searches**

Not only can users search for and display texts from the collection, but they can also search the content of located texts. For example, users can search for Mark Twain's *The Adventures Of Huckleberry Finn*. They can then search the content of The Adventures with a query like "fish and belly" to get a description of Huck Finn's father:

> He was most fifty, and he looked it. His hair was long and tangled and greasy, and hung down, and you could see his eyes shining through like he was behind vines. It was all black, no gray; so was his long, mixed-up whiskers. There warn't no color in his face, where his face showed; it was white; not like another man's white, but a white to make a body sick, a white to make a body's flesh crawl- a tree-toad white, a fish-belly white. As for his clothes- just rags, that was all. He had one ankle resting on 'tother knee; the boot on that foot was busted, and two of his toes stuck through, and he worked them now and then. His hat was laying on the floor; an old black slouch with the top caved in, like a lid.

Moreover, users can search the content of multiple documents simultaneously. For example, users can first locate all the documents in the collection authored by Mark Twain. Next, users can search selected documents for something like slav* (which includes slave, slaves, slavery, etc.). This procedure returns a relevance ranked list of all the paragraphs in the texts containing the word stem "slav".

This feature, the ability to search the content of one or more documents from the collection simultaneously, significantly enhances access to items in the Catalogue. It increases the possibilities for comparing and contrasting themes across texts. There is no reason why users could not locate all the texts written by Thomas Jefferson, Frederick Douglas, Harriet Beecher Stowe, and Abraham Lincoln and then search all of these documents for slav* to begin an analysis the authors' perspectives on slavery.

Since the content search feature supports phrase searching as well as Boolean operations, the feature proves indispensable when users want to put quoted texts into better context. For example, suppose a user is reading a journal article and the author of the article says Shakespeare wrote in *Titus Andronicus*:

> Come, and take choice of all my library,
> And so beguile thy sorrow, till the heavens
> Reveal the damn'd contriver of this deed.

The user can then visit the Alex Catalogue, locate Titus Andronicus, search it for "'my library'", and read the line in its entirety:

> Titus Andronicus: How now, Lavinia! Marcus, what means this? Some book there is that she desires to see. Which is it, girl, of these? Open them, boy. But thou art deeper read, and better skill'd Come, and take choice of all my library, And so beguile thy sorrow, till the heavens Reveal the damn'd contriver of this deed. Why lifts she up her arms in sequence thus?

**Bookcases**

The most unique feature of the system is called a "bookcase". This feature provides the means for users to create customized sets of Alex Catalogue texts for analysis, annotation, and publication. This way users can have their own "study carrel" of frequently used documents.

Users can have as many bookcases as they can remember bookcases names (usernames) and "keys" (passwords). In the event users forget their bookcase names and/or keys, the system includes a means for sending user-defined bookcase name and key hints to previously specified email addresses. Each bookcase can be annotated with text. The text can and is intended to include HTML markup and provides the opportunity to describe the scope of the bookcase. Presently, the length of the annotation is limited by the amount of data sent via the HTTP GET method.

Each bookcase contains one or more "bookshelves". It is on these bookshelves where users save either links to books in

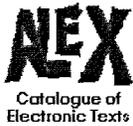

**An Alex Bookcase illustrating a user-defined set of Catalogue items.**

# Create/edit title

Use this page to create or edit a title record.

| Item | Selection | Description |
|---|---|---|
| ID | 26 | Unique identifier |
| Title | Adventures Of Huckle | Enter a type term. |
| Subtitle | | Enter the sub title. |
| Alternate title | | Enter the alternate title. |
| Author(s) | Twain, Mark | Select the author. |
| Author statement | | Enter the author statement. |
| Year conceived | 1885 | Year the work originated. |
| Publisher | Virginia Tech | Select the publisher. |
| Year published | 0 | Enter the author statement. |
| URL | gopher://gopher.vt.e | Enter a type term. |
| Proxy URL | | Enter a proxy URL. |
| Size | 576333 | Enter a type term. |
| MIME type | text/html | Select the MIME type(s) |
| Template type | DOCUMENT | Select a template type. |
| Subject(s) | (No subjects supplied.)<br>Shakespeare, William, 1564-1616<br>text collections | Select the subjects(s) |
| Genre(s) | (No genres supplied.)<br>Allegories<br>Alternative histories<br>Autobiographical fiction<br>Bible plays<br>Bildungsromane<br>test | Select the genre(s) |
| Save and index? | Save and index | Do you want to save and index the file? |
| Re-formatting method | Add blank lines | How should plain text files be re-formatted? |
| Directory | American - 1800-1899 | Where do you want to save files? |
| Note | | Enter any freetext notes. |

[ Edit ] [ Reset ]

Alex Manager version 1.1.7b (06/26/98) © Eric Lease Morgan, 1998
http://sunsite.berkeley.edu/alex/data-entry/index.cgi
Send comments and questions to eric_morgan@ncsu.edu

**A dynamically created Alex Catalogue data-entry screen.**

the Catalogue and/or "bookmarks", the results of content searches. There are no limits to the number of bookshelves any bookcase can contain. Like the bookcase, each bookshelf can be annotated with HTML markup.

To use the bookcase feature, a user first creates a bookcase and/or "unlocks" a previously created bookcase. The user then searches the Catalogue for content, either complete texts or the results of content searches. In either case, the results of searching include links named "Add to bookcase". Selecting these links results in the display of a list. The list includes the names of all the bookshelves in the currently unlocked bookcase. Users are then given the opportunity to select one or more bookshelves where the content will be saved. After returning to their bookcase, the user can continue to annotate the bookcase, bookshelves, as well as the saved links to content.

Finally, the user has the option to "publish" their bookcases. By publishing their bookcases, the bookcase's content becomes accessible but not editable to other readers, and consequently, users can share their analysis and annotations.

Consider the following scenario. A scholar wants to critic all of the Tom Sawyer stories by Mark Twain. The scholar first creates a bookcase and names it something like "Tom Sawyer Studies". The scholar then annotates the bookcase describing its purpose and scope. Next, the scholar creates a bookshelf named, "The Tom Sawyer Books". After searching the Catalogue for all the Tom Sawyer books, the scholar links them to the newly created bookshelf. New bookshelves could then be created representing common themes across the texts. These themes might include the Mississippi River, slavery, coming of age, or friendship. The content of the Tom Sawyer books could be searched. Results of these searches could then be saved in each of respective theme bookshelves. Finally, the bookcase could be published and used as a jumping off point for other scholars or students interested in Tom Sawyer.

### Instant Libraries
Additional access to items in the Catalogue is provided via "instant libraries". Theses libraries are simple compressed archives of all the Catalogue's texts. The archives are divided into the sum of collected American literature, English literature, or Western philosophy texts. Uncompressed, the archives approach 150 of megabytes of data but provide a way for users to download entire subsets of the collection with ease.

Description files (Wide Area Information Server, or WAIS, source files) pointing to the indexes of each collected text as well as two WAIS clients are also available for downloading. By downloading these description files and a WAIS client, a user can take complete advantage of the WAIS protocol and operating system-specific features of the various clients. These clients provide the means for more complex content analysis and even more enhanced access to the texts.

## TECHNICAL INFRASTRUCTURE
The technical infrastructure of the Alex Catalogue of Electronic Texts is really an entire system of files, scripts, programs, and protocols. At first, the system may seem overly complex, but upon closer examination the astute reader will notice many of the items in the system are frequently found in most Web-based information systems. The only truly unique component of the Catalogue's rather baroque architecture is the component named the Alex Manger.

### Operating System
The Catalogue is built on a computer running Unix, specifically Solaris 2.6. Many of the other Catalogue's components run only on Unix computers and therefore the choice of operating systems was predetermined. Incidentally, given adequate hardware, the Catalogue should be able to run on any other version of Unix since none of the system components are specific to particular Unix versions. Not incidentally, the Catalogue is hosted by a fine piece of hardware with more than 1 GB of RAM and more hard disk space than a person can use in a lifetime.

### Database Software
The Catalogue is essentially a database application. The design of the database necessitated multiple tables joined together to form relations. Therefore a relational database supporting SQL (structured query language) made obvious sense. Since Alex is supported by no resources other than use of the hardware and personal time, the database had to be free. A database program called MySQL was evaluated.

MySQL runs under multiple versions of Unix. There are no upper limits for the number of simultaneous users it supports. Unless you sell a product that specifically uses MySQL, there are no license fees. Formal technical support by the developers is provided if you pay for it. Otherwise the Internet community provides technical support. MySQL supports the standard SQL syntax and most of its functions. While MySQL supports only a minimalistic command line interface for the direct manipulation of data, MySQL supports a C and Perl API (application programming interface). For all these reasons, MySQL is the database foundation of the Catalogue. [4]

### Programming Language
An installation of Perl is the next level in the system. The two most important modules included in the Perl installation are DBI (database interface) and LWP, a module supporting popular Internet protocols. Perl is used to directly interact with MySQL as well as provide a foundation for components higher up on the system.

### Cataloging and Classification
The ROADS (Resource Organization and Discover in Subject-based Services) system of software provided the impetuous for migrating the older, gopher-based Alex Catalogue to a more robust system. Based on a simple WHOIS++ template record structure, ROADS provides a complete framework for the creation and maintenance of Internet resource collections. Not only does it provide fundamental features like link checking and record validation, but it also supports controlled vocabularies, Web-based user and administration pages, cross-database searching, and automatic HTML creation. ROADS is one the best kept secrets that shouldn't be. [4]

The Catalogue does not use all the features of ROADS. Specifically, it does not use the input form for record creation, but it does use the ROADS indexer and search engine to provide top-level (metadata) access to items in the Catalogue. Future implementations of the Catalogue will take better advantage of the ROADS automatic creation of subject classification pages.

### Content Searching
A WAIS (Wide Area Information Server, specifically freewais-sf) and a WAIS/HTTP gateway (SFgate) provide the means for searching the content of located texts. Almost every document in the collection is locally archived as a plain text file. If necessary, the files are automatically reformatted so each paragraph (or verse) of the texts are delimited by a blank line. The files are indexed and each paragraph becomes a record of the indexed databases. It is these databases that are queried when searching content. [6]

### Typographically Readable Documents
The documents in the collection are not necessarily intended to be read online. To make the plain text files in the collection more readable when printed, a simple CGI (common gateway interface) script was written based on a shareware utility called txt2pdf by SANFACE Software. [7]

### Data-entry Simplified
The Alex Manager is a single Perl script (with reams of included subroutines) used to facilitate easy data-entry into the Catalogue. Locally developed, it also provides the means for maintaining the controlled vocabularies (subjects and genres) as well as the authority lists (authors, publishers, and time periods) saved in the system's database.

The Alex Manager includes a rudimentary HTTP client written in Perl with the help of the LWP module. Given a URL, the Alex Manager "gets" remote documents. It automatically extracts descriptive information from the documents like titles, dates, MIME types, remote locations, and lengths. New ROADS template records are created through an Alex Manager form containing the extracted descriptive information and pop-up menus of controlled vocabulary terms as well as authority list elements. Selections are made from this dynamically created form. After submitting the form the remote documents are archived locally, indexed with freewais-sf, and finally described with a WHOIS++ template that is automatically generated using ROADS. The result is the acquisition of a new item to the Catalogue with complete indexing and a minimum of human intervention.

The Alex Manager application is also responsible for creating the browsable author and title lists supplementing the Catalogue's accessibility. Without the Alex Manager, the author would not have had the patience to create the Catalogue nor the time to pursue its maintenance.

### Access Control, an HTTP Server, TCP/IP, and HTTP Clients
Basic access control is used to insure the data's integrity against malicious users. An Apache HTTP server is used to facilitate access to the Catalogue. All data is communicated to the outside world via TCP/IP, and anybody who wants to use the Catalogue (except when using a WAIS client) has to use an HTTP server to do so. These are the usual suspects in just about any Web-based information system.

### Policies
While the hardware and software support the Catalogue, the Catalogue is ruled by its collection management policies. These policies govern the purpose of the Catalogue and what it contains. They embody the fundamental principles of librarianship. Namely the collection, organization, storage, dissemination, and evaluation of data, and information for the purposes of expanding knowledge and wisdom. The policies provide a framework so the librarian can do his job and the user can benefit. These policies were outlines at the beginning of this paper.

### Librarian
The librarian is the one who does all the work. She is the one who defines the policies, maintains the collection, evaluates the success of the service, and repeats the process indefinitely.

### Patron
To quote the famous library theoretician, S.R. Ranganathan, "Every reader his or her book." and "Every book its reader." While the technology has changed, the idea behind the quote has not. One of the ultimate ends of the Catalogue is to enhance the learning experience of the user. The Catalogue does not live in isolation and it would not have as much utility, nor would the author have put as much effort into it, unless there were someone to benefit from its use other than himself.

**DISCUSSION**

Since its inception in August of 1998, this new implementation of the Alex Catalogue has consistently received at least 5,300 hits per day. These hits exclude logos and background images. Links from the old implementation of the Catalogue (renamed "Alex Classic") have been redirected to the new implementation. The oldest, gopher-based implementation, despite its lack of maintenance, is still in operation, still receives hits, and is still linked to from around the world.

The biggest users of the Catalogue are people from America Online and domains outside the United States. During the evening, when Internet use in the United States goes down, use of the Catalogue goes up, confirming the use of the system by people outside the United States. Internet spiders are discouraged from indexing the content of Alex through the use of a robots.txt file.

Programmatically, the system is a success since it does not go down, crash, or return misinformation. The system is simple enough that it can be maintained by a single individual. At the same time, the system can accommodate any number of maintainers without sacrificing data integrity. The system is general enough that it can incorporate electronic texts from other disciplines or subject areas with ease.

The cost of the system only includes the initial hardware, $10 (U.S.) dollars for the txt2pdf application, the personal time to the write the Alex Manager, and the acquisition of texts into the collection. This time has been estimated to be around six months of work. Given adequate documentation, this system could easily be adopted by anybody with just about any Unix computer, the articulation of a collection policy, and the time for maintenance.

**FUTURE DIRECTIONS**

In many ways, the Alex Catalogue of Electronic Texts is a success. The system does not go down and the links do not break. At the same time, the Catalogue is not perfect and this last section outlines some of the system's future directions.

In order to provide better intellectual access to items in the collection, the system's metadata should include better analytic access points. In other words, records describing items in the Catalogue should be enhanced with subject and/or genre headings. If this were implemented, then users would be able to browse the collection for things like slavery, love, ethics, mysteries, romances, or satire. The system already has in place a method for adding these headings, but the process simply has not been done.

Another other hand, the scope of the collection could be expanded at the expense of better analytic access. Log files statistically demonstrate users browse the collection by author name more than anything else. This would lead one to believe users are interested in acquiring known items rather than items of a particular subject or genre. If this were true, and by expanding the scope of the documents in the collection to other types of literature besides American, English, and Western philosophy, then the Catalogue might become more useful to more users.

Arguably, the most unique feature of the Catalogue is its ability to allow users to search the content of one or more texts simultaneously. Indexed content accessible through a HTTP/WAIS gateway provides this functionality, but it could be improved. For example, the gateway always returns items in a relevance ranked order. This is not always desirable since a user might want to find the first occurrence of a character in a novel, like Robinson Crusoe's Friday. Additionally, the current implementation of the gateway does not allow the user to navigate between adjacent records; while the gateway returns paragraphs from the texts, it would be beneficial to read the next or previous paragraph in the text without resorting to the complete document.

The bookcase feature allows users to create and annotate sets of Catalogue items for future reference, but its interface is boxy and rather unappealing. In its current implementation, a user's bookcase is intended to be displayed in one window while the Catalogue is displayed in another. Unfortunately, users do not seem to know how to manage multiple windows on their screen. Consequently, the process becomes confusing since they cannot find the appropriate window for editing.

Alex was built with software that is almost 100% free. The only software that incurs a fee is the txt2pdf script, and that fee was only $10 (US). Given more complete documentation describing how to set up an Alex-like catalogue, the system could be compressed into a single archive and given away to others who wanted to create similar collections of texts. Furthermore, if other such collections existed, then it would be technologically possible to search and share texts between collections.

Used effectively, the advent of globally networked computers available to almost anybody in the field of education provides the means for enhancing a student's learning experience. At the same time the process can be quite overwhelming. The Alex Catalogue strives to provide just the right amount of information a user needs for their particular learning experience. Through this process the Catalogue will hopefully stimulate a user's thinking about human values, the processes of art and science, and how all these things can be combined to facilitate a holistic and yet personal method of understanding.

**NOTES**

1. See http://sunsite.berkeley.edu/alex/.

2. See http://sunsite.berkeley.edu/alex/history/announcement.txt.

3. Morgan, Eric Lease. Possible Solutions for Incorporating Digital Information Mediums into Traditional Library Cataloging Services. *Cataloging & Classification Quarterly* 22, 3/4 (1996), Haworth Press, pp. 143-170, and Craigmile, Bob and Wohrley, Andrew. Searching Cyberspace in Ensor, Pat, ed. *The Cybrarian's Manual.* American Library Association, Chicago, 1997. pp. 155-6.

4. See http://www.tcx.se/.

5. See http://www.ilrt.bris.ac.uk/roads/.

6. See ftp://ftp.germany.eu.net/pub/infosystems/wais/Unido-LS6/freeWAIS-sf/ and
http://sunsite.berkeley.edu/~emorgan/SFgate/SFgate.html.

7. See http://www.geocities.com/CapeCanaveral/Hangar/4794/txt2pdf.html.